# VISHIEN-MAAT: Scrollytelling Visualization Design for Explaining Siamese Neural Network Concept to Non-Technical Users

Noptanit Chotisarn[a], Sarun Gulyanon[c], Tianye Zhang[a], Wei Chen[a,b,*]

[a]*The State Key Lab of CAD&CG, Zhejiang University, Hangzhou, China*
[b]*Laboratory of Art and Archaeology Image, Zhejiang University, Ministry of Education, China*
[c]*College of Interdisciplinary Studies, Thammasat University, Bangkok, Thailand*

**Abstract**

The past decade has witnessed rapid progress in AI research since the breakthrough in deep learning. AI technology has been applied in almost every field; therefore, technical and non-technical end-users must understand these technologies to exploit them. However existing materials are designed for experts, but non-technical users need appealing materials that deliver complex ideas in easy-to-follow steps. One notable tool that fits such a profile is scrollytelling, an approach to storytelling that provides readers with a natural and rich experience at the reader's pace, along with in-depth interactive explanations of complex concepts. Hence, this work proposes a novel visualization design for creating a scrollytelling that can effectively explain an AI concept to non-technical users. As a demonstration of our design, we created a scrollytelling to explain the Siamese Neural Network for the visual similarity matching problem. Our approach helps create a visualization valuable for a short-timeline situation like a sales pitch. The results show that the visualization based on our novel design helps improve non-technical users' perception and machine learning concept knowledge acquisition compared to traditional materials like online articles.

*Keywords:* Story synthesis, Scrollytelling, Visual storytelling, Visualizing deep learning, Learning science

## 1. Introduction

In the modern world, artificial intelligence (AI) is important and plays an important role in business, society, and everyday life. The main obstacle in AI adoption is that AI usually comes in the flavor of a complex and black-box solution, which makes users skeptical since it is difficult to understand its concept, process, and task. Hence, building trust in AI is key to successful AI adoption, and that is what explainable AI (XAI) aims to achieve through improving the transparency and understanding of machine learning (ML) models to create a safe and trustworthy ML model [13]. However, before XAI can be applied, one question arises: how does the ML model work and obtain the solution? This particular question is typically asked by customers, which are non-technical stakeholders of AI-related products. During the sales meeting, the business marketing team, which normally are non-technical users, struggles with this question because they have to convey the complex idea within the time limits, and the publicly available materials, like online articles and videos, are either too abstract to follow or too academic to comprehend. In such a situation, the material with the ability to control the pace of the narrative and details of the content is of utmost importance; meanwhile, the material also has to be intriguing and appealing to the users and break down complex ideas into easy-to-follow steps.

One possible solution is storytelling visualization, a well-known and influential means of communicating messages and engaging audiences, suitable for non-technical users. One popular technique, called 'scrollytelling', has recently gained traction as it engages readers more naturally and richly by unfolding an expressive visual story depending on their interactive inputs. It shows the changes in a website's content as users scroll through the page. The content usually consists of visual graphs with associated narrative writing, video/audio clips, and interactions [25]. Because scrollytelling gives the users complete control of both the pace and the details of the visualization, it is a versatile tool for explaining complex content to various audiences, especially for non-technical stakeholders. These properties make scrollytelling a practical tool for the business marketing team's situation to help customers better grasp and absorb the knowledge about AI concepts.

One caveat in using effective scrollytelling is that the visualization design must align with the content; otherwise, it may lead to faulty interpretation, overwhelming visualization, and/or user frustration. To avoid these problems, our work adopted two main concepts in formulating the scrollytelling and its interaction components: a) the scrollytelling visualization design concept and b) the configuration in-between the scrolling concept. The first concept ensures that the content is laid out so that the scrollytelling can be applied effectively. While the latter concerns how to make the interchange between

*Corresponding author
Email addresses: chotisarn@zju.edu.cn (Noptanit Chotisarn), sarung@tu.ac.th (Sarun Gulyanon), zhangtianye1026@zju.edu.cn (Tianye Zhang), chenvis@zju.edu.cn (Wei Chen)

the fine- and coarse-grained levels of description noticeable and intuitive.

To demonstrate our scrollytelling visualization design, the visual similarity matching problem is selected as it has many applications, e.g., visual recommendation and visual search; and it is the task assigned to the business marketing team, where we conducted our experiments. One of the popular techniques for the similarity matching task is the Siamese Neural Network (SNN), which has been implemented in the AI solution for real-world business problems [7]. In a sales pitch, the business marketing team normally gives a presentation and demo of the AI product, which only introduces the front end of the product. The team usually has trouble when there are questions about the back end, so the team needs an appealing tool, like the scrollytelling following our visualization design, to help introduce the SNN technique to the funders and management executives who are not familiar with the AI concept. Due to business confidential information and to make this work user-friendly, the scrollytelling visualization illustrates the SNN concept through cat breed images instead of product designs for which is intended.

To verify the utility of our approach from the non-technical users' point of view, the observational study and feedback from the business marketing team are presented. Another important question is how the scrollytelling, created by using our visualization design, compared against other mediums in terms of ML knowledge acquisition for non-technical users. Due to business restrictions, we can't evaluate the executives or investors with the full questionnaire. Instead, we conducted experiments on another group of non-technical users, business IT students with no background in AI. We chose this user group because both investor and student users are on the receiving end, and both are unfamiliar with the SNN concepts. Student participants were asked to use different mediums, such as scrollytelling and online articles, and took the tests to assess their ability in knowledge acquisition.

The contributions of this work include:

- A novel scrollytelling visualization design for explaining the SNN concept in the visual similarity matching problem context from real business scenarios.
- The visualization design effectively improves the understanding of non-technical users in how the deep learning model works.
- A user study compares the ML knowledge acquisition of non-technical end-users across different mediums such as scrollytelling and online articles.

The structure of this paper is as follows: a review of relevant literature in Section 2 and the explanation of the SNN model used to demonstrate our visualization design in Section 3. Then, the novel visualization design of the interactive scrollytelling is presented in Section 4. The evaluation and the case studies of the visualization are explained in Section 5. Section 6 shows the discussion of the aspects of visualization that result in the achievement of learning ML model comprehension. Finally, Section 7 summarizes the conclusions.

## 2. Related Work

*2.1. Visual Storytelling*

A visual narrative, also known as visual storytelling, is a story told primarily using visual media, which makes it appealing to the audience and easy to follow. The story can be told through still photography, illustration, or video and supplemented with graphics, music, voice, and other audio. A visual narrative is any story told visually [3].

Two issues must be addressed to create an effective visual narrative: the conversion of data into a story for communication purposes and narrative visualization. One of the methodologies for bridging the gap between gathering data and communicating is *story synthesis* [4]. Story synthesis provides an easy-to-use framework for making sense of complex data and attempts to assist stakeholders in turning analytical results into actionable information. To tell a story out of the findings, the analyst must first describe them as *story slices*, which are structured information derived from the study's original data. The story slices are then arranged in the appropriate sequence to reveal the connections between the various pieces of data. As support for implementing the visual analytics system, the story slices should include system functions. As a result, through data-driven story slices and narratives built from synthesized content, the story synthesis enables two-way linkages between the phases of research and storytelling, which can apply the story synthesis technique for data journalism [6].

*2.2. Scrollytelling*

Creating a logical series of related data-driven visualizations, or visual pieces, required to present a message engagingly and successfully is known as *narrative visualization* [18]. On the web, scrollytelling is a visual storytelling approach for narrative visualization. Scrollytelling, also known as explorable explanations, dynamically updates the contents of a website when web page viewers scroll the page. Scrollytelling provides readers with a natural and rich experience. In short, the term 'scrollytelling' was coined to characterize long-form internet storytelling that incorporates audio, video, and animation [24, 14].

Scrollytelling comes in a variety of forms, depending on the effect that scrolling a web page has as follows: *Scroll As Steps*, *Continuous Scrolling*, *Scroll As A Trigger*, *Mixed Scrollytelling* [1]. Readers may not have to guess what to tap, click, or swipe to engage the tale. A user's position could also trigger multimedia events like video playback, animation, and transitions—a dynamic interplay of text, visuals, and music. In addition, the study in [21] shows that scrollytelling and video provide much more memory than audio and, to a lesser extent, text media, which is suitable for non-technical users.

Currently, little research has been conducted on scrollytelling to demonstrate how a machine learning model works [24]; it may be found in the Distill[1], a collection of work on describing machine learning in web scrollytelling.

---
[1]https://distill.pub/

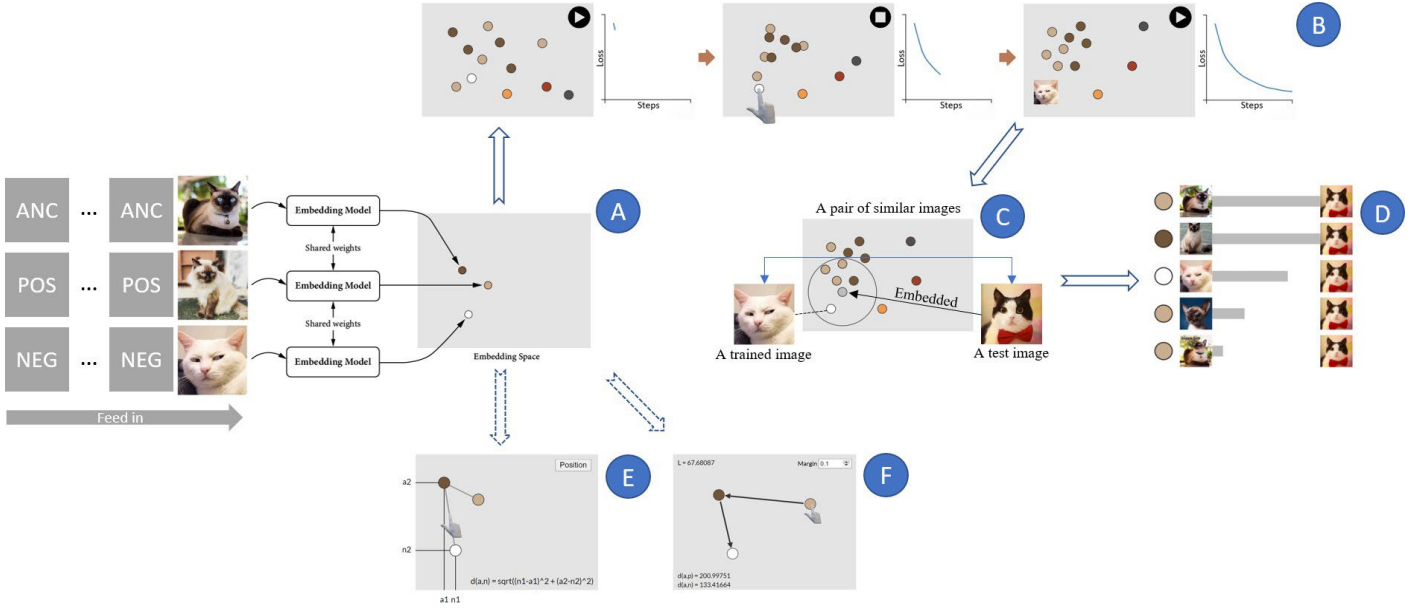

Figure 1: The embedding model (A) describes how image batches are fed into the feature-matching model and plots images as vectors in the embedding space. The solid arrow's path (from A-B-C-D) depicts the core path of the model training step to get the trained model (B). Then inferencing to find the nearest neighbor of a test image among the trained images, which are bubbles inside the circle (C), and present the inferencing results from the previous bubbles and place them as the bars of similarity distance with the test image (D). The dashed arrow's path provides related concepts that the users have to know regarding calculating the Euclidean distance to detect similarity (E) and the triplet loss function used during training (F).

## 2.3. Mediums for Storytelling

Apart from scrollytelling, there are other mediums for narrative visualization. In Table 1, we compare different mediums, including online articles, videos, data-GIFs, data comics, and scrollytelling. The comparison is based on the supported content types, such as text, image, and animation, and the available features, such as play/pause, skipping, scrolling, and interaction.

The articles are published in a newspaper, magazine, and, most notably, online articles. Authors can convey many messages through text and illustrations, whereas online articles offer more interaction options than paper-based.

Videos and video clips are used broadly to refer to any video program uploaded to a website or other medium. Content creators can tell stories with full animation and graphics. Viewers can skip, rewind, forward, and pause the video but cannot interact with it.

Data-GIFs are a new type of data-driven storytelling that employs simple visual messages embedded in 15-second animations. Creators can include as many graphics and animations as they want, but they have less narrative time. The interaction options with GIFs are limited as they can only be played or paused.

Data comics are a novel approach to data-driven storytelling that employs sequential art inspired by the visual language of comics. The reader will feel more relaxed, similar to writing on-demand long-form online articles, but emphasizing images rather than text. Nonetheless, readers are unable to interact with the comics.

Compared to these mediums, scrollytelling supports all content types and has the most features. In addition, it also

Table 1: Comparison of ability and characteristics between different types of mediums

| Mediums | Text | Image | Animation | Play/Pause | Skipping | Scrolling | Interaction |
|---|---|---|---|---|---|---|---|
| Articles | ✓ | ✓ | ✓ | * | * | ✓ | * |
| Video | ✓ | ✓ | ✓ | ✓ | ✓ | - | - |
| Data-GIFs | ✓ | - | ✓ | ✓ | - | - | - |
| Data Comics | ✓ | ✓ | ✓ | * | * | ✓ | * |
| Scrollytelling | ✓ | ✓ | ✓ | * | ✓ | ✓ | ✓ |

An asterisk indicates the presence or absence of the features.

gives users the flexibility to control the pace and provides a frictionless way to digest content. Thus, we select scrollytelling as the medium for our visual storytelling.

## 3. Siamese Neural Network

This section explains the SNN used to showcase the proposed scrollytelling visualization design. First, the dataset used for demonstration is described followed by the key concept of SNN like distance metric learning (DML) and important components of SNN.

## 3.1. Cat Breeds Dataset

To make this work user-friendly, lovely cat pictures have been scientifically demonstrated to increase caring and concentration [17]; The dataset of cat breeds[2] was chosen (Figure 2) to represent the interaction components via this proposed scrollytelling visualization for the internal model training process for non-technical consumers to notice the concept of a deep learning classifier for product design matching.

---
[2]https://www.kaggle.com/datasets/ma7555/cat-breeds-dataset

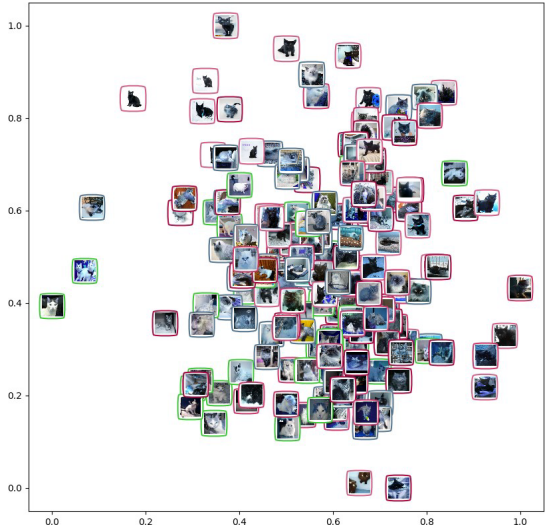

Figure 2: The cat breeds dataset visualized using t-SNE.

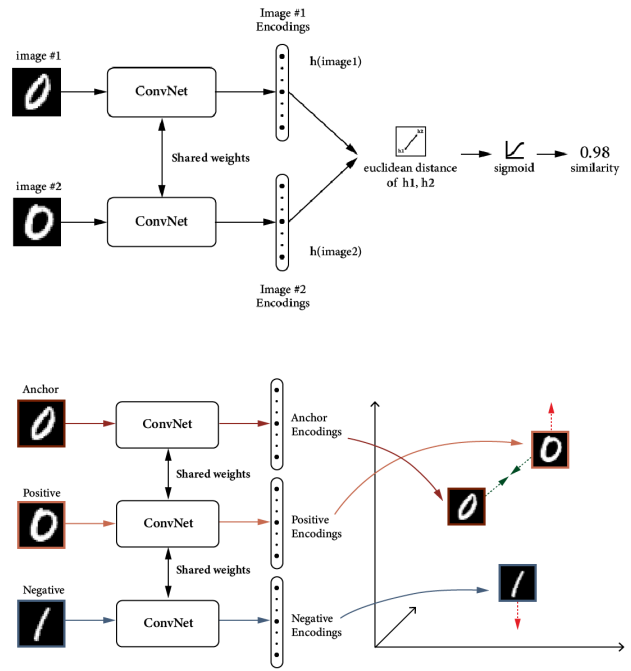

Figure 3: The architectures of the SNN model with two different loss functions: contrastive loss function (top) and triplet loss function (bottom).

We adopted the cat breeds dataset in our scrollytelling visualization design as a neutral dataset to be used anywhere and for any purpose in the presentation instead of using real products from the real business directly. Moreover, the cat dataset was related to the system's name, inspired by a Google image search for 'Siamese,' which returned the 'Siamese cat' known in Thai as 'Wichienmaat,' which we pick up as a homophone name for the system as 'VISHIEN-MAAT,' where 'VIS' stands for visualization.

### 3.2. Distance Metric Learning

DML aims to learn a transformation that converts images into a representation space where distance corresponds with a notion of similarity [19]. Example applications of metric learning include zero-shot learning [16, 5], visualization of high-dimensional data [15], dimensionality reduction [11], and face recognition and clustering [22]. There are several interactive online articles explaining how DML works and how it can be applied, like 'How to Use t-SNE Effectively' [29] that allows users to experiment with the parameters of the t-SNE algorithm, and 'Visualizing MNIST'[3] that visualizes different types of embeddings produced by different algorithms.

A technique for DML is the SNN [2], which is a sort of neural network that consists of many instances of the same model with the same architecture and weights. The embedding models for the query and reference images are learned simultaneously using the standard parameters. The vector representations of the two images are then compared to determine how similar they are. For both feature extraction and metric learning, SNN is a very powerful architecture [9]. An example of previous work in visual similarity matching using SNN and convolutional neural network (CNN) is the method in [27] to learn dyadic item co-occurrences. Another example is the method called 'MILDNet' [28], which is the SNN that uses a single CNN with skip connections as the subnetwork to encode both query and reference images as vectors; which, in turn, they are used in the calculation of similarity distance.

### 3.3. Embedding Model

Pixel images are more than raster images containing no direct information about the shape or structure of the objects in the image—the embedding model stores images in embeddings. In this low-dimensional space, crucial features for similarity matching are recorded. The embedding model is commonly computed using a deep convolutional neural network (CNN or ConvNet), which turns input images into embedding vectors.

The model's back end is a standard multilayer perceptron layer to interpret the CNN features. A convolutional layer, a pooling layer, and a fully connected layer are the three layers that make up a CNN. A conservative CNN configuration uses filters and a kernel with an activation function. This is followed by a pooling layer, which reduces the convolutional layer's output. The CNN output is then flattened into one long vector to represent the 'features' extracted by the CNN, which we call the embedding vector. To see how these vectors are grouped together, we use t-SNE to show the vector representations from embedding models. The semantic comparison can only be made where the distance between vectors represents the similarity between objects.

---
[3]https://colah.github.io/posts/2014-10-Visualizing-MNIST/

*3.4. Architecture*

SNN is a type of neural network that contains multiple instances of the same model with the same architecture and weights (Figure 3). It is an exceptionally robust architecture for feature extraction and metric learning [2, 9]. The embedding models for the query and reference images are learned simultaneously using the standard parameters. The vector representations of the two images are then compared to determine how similar they are.

*3.5. Loss Functions*

SNN is optimized based on the similarity between the query and the reference images, which the loss functions can measure. The two most prevalent loss functions for SNN are the contrastive [5] and triplet [22] loss functions (Figure 3). The contrastive loss is a distance-based loss for learning the embeddings that give similar embedding vectors for positive sample pairs and different embedding vectors for negative sample pairs. On the other hand, the triplet loss necessitates three inputs: baseline (anchor), positive sample, and negative sample. It looks for the embeddings that shorten the distance between the anchor and the positive sample while lengthening the distance between the anchor and the negative sample.

*3.6. Vector Representations Visualization*

The vector representations learned by the SNN model can be visualized to identify how these embedding vectors are clustered together using the t-distributed stochastic neighbor embedding (t-SNE) technique, a non-linear dimensionality reduction approach for mapping high-dimensional vectors to a lower-dimensional space (*e.g.*, 2D or 3D) for data visualization. The embedding vectors of the cat breeds dataset are displayed using t-SNE in Figure 2, where the object's category colors the data points. In the same manner, similar design styles should be mapped to similar vectors and clusters in the product design matching problem. In contrast, samples with mismatched design styles should be mapped to different vectors and clusters. Even though we were inspired by t-SNE, which is used in our ML project from real business, other dimensionality reduction techniques can still be applied instead of t-SNE. So, the embedding model and vector representations visualization can be considered different dimensionality reduction techniques, not only t-SNE.

## 4. Visualization Design

To fulfill the requirements of both the business marketing team and the business IT students, first, we formulate the design rationales for creating a scrollytelling visualization that can introduce the SNN concept in an attractive and user-friendly way to non-experts in the sales pitch scenario. Here, we explain how to use our visualization design and discuss the interaction components, which are the most important visualization parts when explaining complicated ideas.

*4.1. Design Rationales*

When explaining complex AI concepts to non-technical users in a short-timeline situation like a sales pitch, the presentation materials should be appealing but not overly technical or academic to communicate with the general audiences, *i.e.*, non-technical users. Most learning materials about the AI concept, particularly SNN, are in the form of text in online articles. These materials require learners to be somewhat familiar with the fundamental concepts or do not break down the materials sufficiently for non-experts to understand. Storytelling aims to convey only relevant and essential information, and the information should be presented in a straightforward and easily accessible manner. These reasons make storytelling suitable for our case.

Scrollytelling is an effective method for storytelling with more dynamic and friendly interactions for non-technical audiences who need quick, precise, interactable, and easy information without needing advanced skills. It blends text, photographs, graphics, video effects, and animations that flow seamlessly to create exciting tales and inventive approaches for displaying a story or information flow better than other methods, such as infographics, data comics, or data-GIFs. In addition, the ability to control the pace and adjust the levels of description are useful for these audiences because they may want to look back at some details or find out more about particular points, where we can add extra interactions that help break down the complex ideas even further.

To create an effective scrollytelling for storytelling, we use two main concepts in formulating the scrollytelling and its interaction components:

1. *The scrollytelling visualization design concept* serves to ensure that the storyline aligns with the nature of scrollytelling, where the story progresses in a straight line, and users scroll through the content. In our case, this concept gives users control over the narrative of the SNN concept, where users may scroll down and peruse the content by hovering over the selected items in the tale components, such as text descriptions next to the graphic elements, to highlight the information. As a result, this concept gives users complete control of the narrative pace.

2. *The configuration in-between scrolling concept* emphasizes the characteristics of interaction components of scrollytelling that must complement the storytelling for explaining complex concepts, and they must be noticeable and intuitive. In our showcase, this concept helps depict the changes during the interchange between the fine- and coarse-grained levels of description for the SNN operation. It helps explain complex ideas, such as parameter tuning and optimization, during model training. Consequently, users gain a better understanding of the tuning equation by analyzing the interplay of the visual elements in the tale components.

The steps for creating the scrollytelling visualization design include analysis and storytelling. We collect and organize information components, choose the content to be displayed, and assist in creating stories [23, 8]. Story synthesis is used to help handle information components by converting raw analysis findings into story slices that are ordered based on the critical linkages between them. It created a story that successfully conveys the operation of the machine learning model to recipients by utilizing appropriately formatted story slices to aid in story creation [26, 4]. We define the system functions required to explain the AI concept and to generate story slices ranging from showing visual displays, handling the collected objects, and supporting the story synthesis activities.

*4.2. Applying the Story Synthesis Framework*

The generic conceptual framework defines story synthesis as the process of generating story content and structure. This framework can be used in designing visual analytics systems that provide support for story synthesis, as detailed below:

*4.2.1. Define the types and structures of story slices*

The story creator envisions the facts or patterns that can be identified and used as story slices based on the data and the analytic aims. Machine learning studies are often divided into sub-topics according to each model: input, feature extraction, classification, and output. Those sub-topics can be described based on the purpose of the specific use. For the SNN, we define the types and structures of story slices into six topics that we will use for different designs; the six-story slices are started based on an overview to show the concept of SNN, the related concept, *e.g.*, embedding model, and euclidean distance. The stages of the model, *e.g.*, loss function and model training and inferencing of the model results. To describe the aspect that the user needs to understand as follows:

1. **SNN Concept** slice explains the SNN, its origins, and its nomenclature. Users should grasp the model concept created from the twin convolutional neural network (CNN).

2. **Embedding Model** slice explains how to convert an image to a vector using an embedding model with a CNN background. Users should comprehend the concept of using CNN to turn a batch of images into embedding vectors placed in space.

3. **Euclidean Distance** slice describes the calculation used to determine similarity. Users should grasp the relationship between transformed vectors put in space that can be utilized to calculate euclidean distance equations. In subsequent phases, the distance results can be used to determine the loss function.

4. **Loss Function** slice demonstrates how the model's performance changes when different functions are used to calculate it. Users should comprehend when the distance changes concurrently with the movement of the embedding vectors.

5. **Training** slice depicts the training process's metamorphosis. Users need to watch the automated animation of the training process and understand what is going on in the background when the model is being trained in the development environment.

6. **Inferencing** slice represents the process of using model inference. Users should understand how the model's output can be used. One method is to find the similarity of a new image to the images in the trained embedding space.

The last five slices will be treated as interactive components except the first one, which is the image.

*4.2.2. Design a representation for a story slice*

After all six story slices have been drawn, the next step is to analyze what kind of presentation would be appropriate for the narrative of each story slice. The story creator chooses a suitable data structure, such as a graph or a vector, to represent extracted narrative slices internally in the system based on the story slice structure. This depiction may or may not be included in the final story; as a result, the story creator may decide to replace it with something else that is more appealing to the recipient during the storytelling stage. Fact types, visual graphs, and annotations are three areas in which a work [14] outlines rules that can build a set of visualization sequences as scrollytelling stories.

The image and image description will be used to convey the story, and their interactions will be used to clarify further the story slices of the SNN concept (*e.g.*, Embedding Model, Euclidean Distance, Loss Function, Training, Inferencing). The scatter-like bubbles were chosen to represent the import image data processed by the SNN. An interactive bubble chart was used to depict the model's story slice because the image data will be translated into space by placing a scatter-like containing image data. Bubbles' distribution was dispersed throughout the rectangular space. For bubbles' color encoding shows the difference and association between attributes. Our work also represents the color palette as brown, gray, black, and white for color encoding, which comes from the color palette of the Siamese Cat.

*4.2.3. Define story synthesis support functions*

The next step is to analyze what functions the bubbles will have. The system functions are required for creating story slices or extracting story slices from visual displays, managing the objects retrieved, and supporting story synthesis activities. The SNN model concept was used to narrate the story slices. The bubbles within each story slice, also known as interactive components, have their own corresponding function, that can be interacted with by the users. For each interactive component, the corresponding functions as narrate as a story was created that may be difficult to understand for non-technical users, such as a math equation-like component, so it has more explorable by self-interaction and scrollytelling.

*4.2.4. Design the visual analytics system*

The functions have been developed with bubbles into a visual analytics system. It must analyze the system based on the user and the techniques used to represent the support functions. This includes creating interactive and manipulable tools for the reader to find facts or patterns that may become story slices, such as clickable buttons or animated images that unfold for the reader as they scroll through the story. The scrollytelling visualization [10] will be used to depict all of the defining and designing of story slices from the story synthesis framework, allowing non-technical users, *e.g.*, students who have never learned machine learning as a prerequisite course, to analyze the application of the SNN and recognize knowledge from the visualization [12].

*4.3. Interaction Components*

Stories were presented using the six interaction components. These are the components where the user will pause during the scroll to understand the corresponding functions narrated as a story through each component better. It uses ReactJS in conjunction with the VEV[4] platform on the front end. The backend manages data, uses Python to train and test models, and then generates JSON for the front end to render.

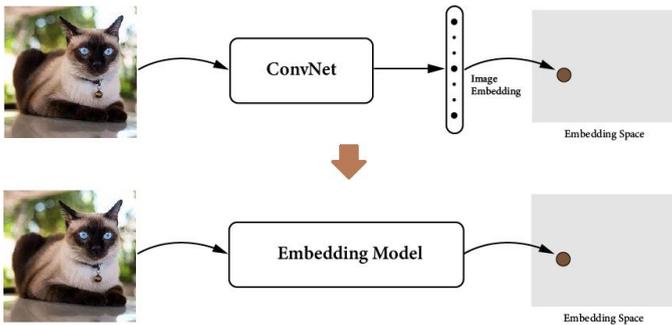

Figure 4: Hovering over the image reveals the embedding model's internal operations.

*4.3.1. Showing encapsulation of the embedding model*

One proper approach to show the fast state transitions of non-complicated components during scrolling is to place the mouse cursor across that component and change the state to 'image hover'. That frees the user from thinking more about what this component wants to represent. The user can understand as soon as his mouse moves and hovers, intentionally and unintentionally.

The first component of interaction is Image Hover effects enable the addition of interactivity to elements on a website without slowing it down. Hover effects are elegant, do not clog designs, and help websites run smoothly. The Image Hover (Figure 4) was used to demonstrate the second story slice, Embedding Model, which is the core for image embedding that encapsulates the CNN or ConvNet, the sourcing process for positioning an image into an embedded space.

---

[4]https://a-vishienmaat.vev.site/siamese

*4.3.2. Describing transition state of the space embedding*

To show the flow of data while changing the state of the data through different processes, it is possible to use a comparative representation of how the previous data was flowing through the process and the final result via 'image compare'. To scroll down to this point, the user will have to pause longer because there are many parts to communicate with this component, including snippets, processes, and results. Providing users with a pre-and post-process of state transition of space embedding which comparison can make them more understandable.

Image Compare is the second interaction component, a simple but fully customizable before/after image comparison component. The component adds a vertical or horizontal slider to two overlapping images, allowing the user to adjust the slider to examine their differences. For a basic description of batch feeding to the embedding model for the process of locating the images for each batch into the embedded space, used the Image Compare (Figure 5) to support first and second story slices.

*4.3.3. Variables of distance equation*

The commonly used method to show the data relationship in a node graph is to hover over a node and its node link called 'line hover'; here, the nodes are represented by the bubbles and the link between the nodes is the origin of the distance equation. When scrolling down to the component hovering over the node link, this is forth the user stops considering nodes and their relationship.

They were hovering over a line expressing data functions similarly to hovering over an image. This will indicate how the value of that line has changed. Hovering on lines is used in this work to demonstrate the difference in how the distance between each line is calculated and to describe the distance between bubbles. A bubble is a point in an embedded image where a line is generated by joining two bubbles into two pairs (Figure 6) for supporting the third story slice. The first pair includes the anchor image's position and the image closest to it, called the positive image. The second pair consists of the anchor image and the image that differs from it, called the negative image. The similarity or differences may be calculated from the Euclidean distance equation.

*4.3.4. Interpreting distance computation*

In addition to showing the correlation of data in a node graph by hovering over a node and a node link, each node is constantly changing shape. This is normal for model training, so the relationship between nodes will also change. Scroll down to this component to manually simulate the change in the node's position, which is the user's bubble via the 'draggable component'. It allows the user to freely change the position of the bubbles by dragging and dropping them to different places and seeing the change in the distance of each bubble by interpreting the distance and loss computation. When dragging an object to a new position, the distance between all objects is calculated and displayed as a loss value. This allows viewers to understand if objects of similar color are closer, the loss will decrease. Users can try iterating several times manually to see how the drag-and-drop changes in new locations.

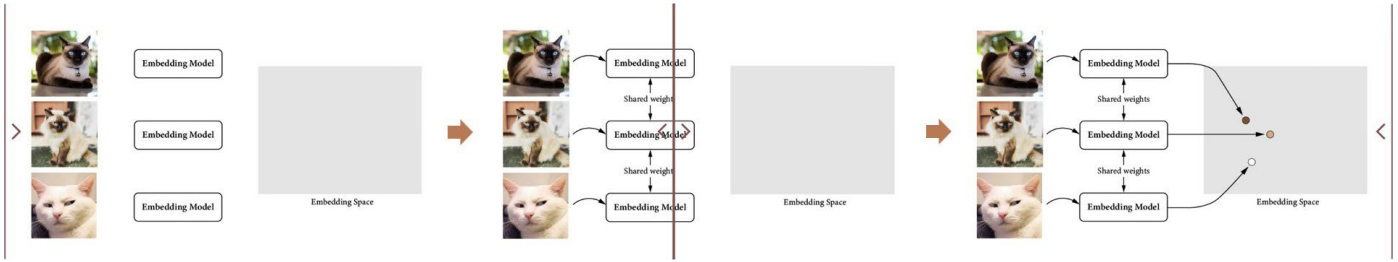

Figure 5: Image comparison by sliding the bar left/right reveals the images' transformation into bubbles on the embedded space.

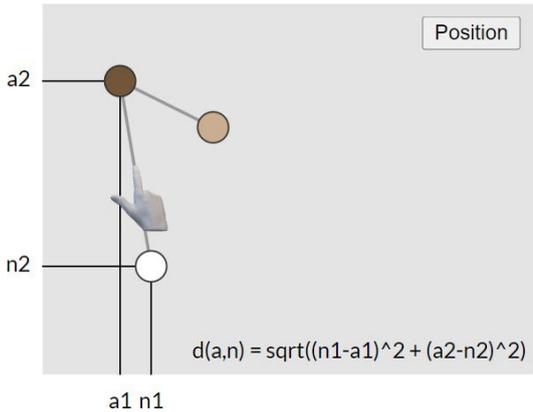

Figure 6: The variables of the Euclidean distance equation are described by hovering the lines between the bubbles.

The draggable functionality on any DOM element allows moving the draggable object by clicking on it with the mouse and dragging it anywhere within the viewport. The draggable component (Figure 7) was used to help users understand how to calculate the loss value manually by dragging the bubbles representing the location of the embedded images, which is the loss function story slice. With each training cycle, the position of the bubbles will change according to these situations, the position of the positive bubble will move closer to the anchor bubble, and the negative bubble will move away from the anchor bubble. The user can simulate the above situation by dragging the three bubbles closer to or farther from each other. The closer the positive bubble moves to the anchor bubble, the greater the loss decreases. The further the negative bubble moves away from the anchor bubble, the greater the loss decreases.

*4.3.5. Showing the model training step*

To show the model training changes in each epoch, a continuous motion picture of bubbles is gradually grouped; this section is highlighted, showing the grouping of dispersed bubbles. However, the change in each epoch happens quickly, so to see the movement in each step, it is necessary to pause. The user can understand each bubble by hovering the mouse over the bubble's location of interest. The properties we can use for this need are 'animation-play-state', in which the user can pause/play the stage-change animation in CSS.

The animation-play-state sets whether an animation is running or paused. Resuming a paused animation will pick up where it left off when it was paused, rather than starting from the beginning of the animation sequence. The image in the actual training model is a moving image (replaced by bubbles) during the development process. Under the initial condition, similar images move closer, and different images will move away, which can calculate the loss. The loss value is calculated for each training cycle. It is plotted as a graph between the x-axis (number of training cycles or steps/epochs) and the y-axis (loss), where the expected loss will be decreased gradually for each step. This loss chart is used to decide when to stop training; when the loss chart decreases and runs as a plateau for a while, the developer will order the training to stop. That means no further training has been learned, and the loss has not decreased significantly anymore.

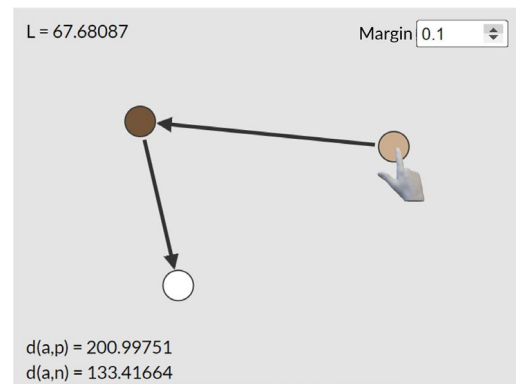

Figure 7: Drag and drop bubbles in various spots and set the margin to see the loss calculation.

The processes can also be displayed as animation-play-state (Figure 8), so the user can press play/pause to view the model's training on-demand and hover over each bubble to view the original image of each bubble, which is the training story slice. It will find that the bubbles are encoded with the primary color of the cats, *e.g.*, black for black cats and orange for Ginger cats. The similar colors (dark brown and light brown stand for Siamese Cat and Balinese Cat, respectively) will move towards each other, which means it is a similar image. Both breeds of cats are similar in color but differ in hair length. Moreover, bubbles with different colors from those, such as orange and black, will move away from the group of dark brown and light brown bubbles. This means they are separated by a distinctly different color to the brown group (Siamese and Balinese), for example, the Ginger cat, black cats, etc.

*4.3.6. Inference a new embedded image*

Allowing users to interact with the process of implementing the model in real-world scenarios: importing a new image, which is the inference of the new image as embedded in the system to compile it with a trained model. Importing corresponds to different ways of communication, such as uploading, putting, and throwing. Directly presenting the colloquial language is better for communication. So we chose to use 'drag and drop,' representing putting or throwing the new image into the space of the trained model.

In a complex drag-and-drop interface, while keeping the components decoupled, the components change their appearance and the application state in response to the drag-and-drop events. It is a perfect fit for dragging data transfers between different application parts—the inferencing section, which represents the state of transition of an image for testing with the model. The test image will be thrown into the Embedding Model, and then it will automatically be embedded and shown into the space. Drag and drop were chosen in this throwing because it is an easy-to-understand interaction for the user by having the user interaction initiate the flow of inferencing (Figure 9), which is the sixth story slice.

In addition, a simple interaction, such as clicking on a newly changed point, was applied to the test image that had just been embedded in the space. The test image was replaced with a bubble and encoded with a new color that was distinct from the other bubbles in the space. Users can view similar members of the test bubble by clicking on the newly embedded bubble; the system will display a circle centered on the test bubble with a radius covering the nearest bubble side by side. Moreover, the system will show the nearby image by comparing the distance. Finding the closest object can be calculated using an algorithm like the k-nearest neighbor.

## 5. Evaluation

We began by describing the two scenarios in which our scrollytelling visualization design has been used: for business presentations and learning the model. We use observational and user studies to get a qualitative and quantitative look at the two use cases. The study's structure has a participant, procedure, and feedback or results.

*5.1. Usage Scenarios*

Two scenarios explain the scenario used for the business marketing team's business presentation in a company and student learning of the business IT student who had never learned about machine learning.

*5.1.1. Business presentation scenario*

This scrollytelling was created as one of the SNN model's presentations as part of a project for an image-searchable e-commerce website. The SNN model groups images of building materials and upholstered furniture into clusters. The model builder's issues are that they must modify the presentation each time they meet a new group audience to make it relevant. Because they have to remember terms for building materials and architecture as well as the complex ideas of the deep learning model, it can be hard for the audience to understand how the model works when it is shown with real-life product pictures.

Cat images were used to explain how the model works to make it easier to remember domain-specific terms. Audiences who are non-technical do not need to pay attention to the domain terminology because cat images are easy to get familiar with. In addition, to make it easier for an audience to digest, the complex content of the deep learning model is broken down into parts so the narrator can describe the model in sections by arranging story slices according to the model's training process. Some things, like math equations, may be hard to understand while listening to the story, but they can be made clear by playing with the bubbles that show the equations. Users can play through the scrollytelling website at the audience's pace or be used in the presentation at the presenter's pace, making our scrollytelling visualization design for explaining deep learning concepts to the audience. The three types of user results will be shown in the observational study in Section 5.2.

*5.1.2. Student learning scenario*

Knowledge about the algorithms of various deep learning models is necessary for the management information system (MIS) nowadays. People working in this area in the MIS field need to understand various business requirements from technical and business stakeholders since MIS people are in the middle between the very technical developers and the non-technical business users. Nevertheless, teaching deep learning concepts to business IT students who learn about MIS can be tricky since they need to understand the complexity of the concepts from both points of view—technical and business. This scrollytelling work can be an effective teaching tool for students, where student users can learn on-demand by scrolling and reviewing the interactions in each story slice representing each model process which is an alternative to online articles. The results of this user group will be shown in the user study in Section 5.3.

*5.2. Observational Study*

We conducted an observational study to investigate how VISHIEN-MAAT's target users (*i.e.*, the business marketing and related teams in a machine learning (ML) project) would use this scrollytelling to learn about SNN and give its feedback.

*5.2.1. Participants*

We asked the presenters from the business marketing team from real business about their experiences with the scrollytelling that explains the SNN model to evaluate the scrollytelling visualization that illustrates the AI concept in the ML project related to SNN. They have been working on parts of this ML project for about three years. Also, the customer's audience's opinions and feelings are gathered during the meeting.

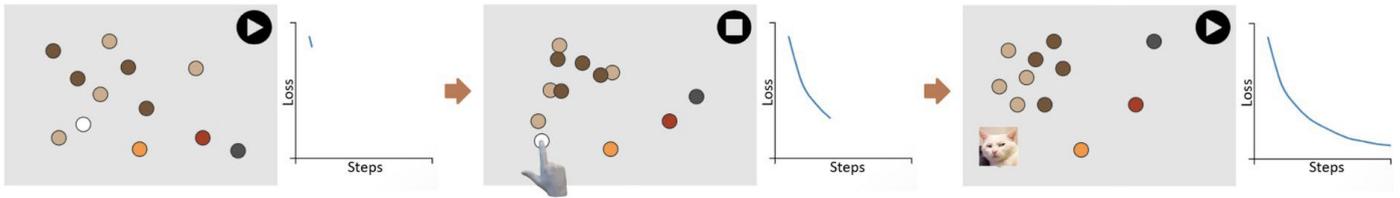

Figure 8: The user may observe a training model that compares bubble movements to loss graphs by clicking the play/pause buttons.

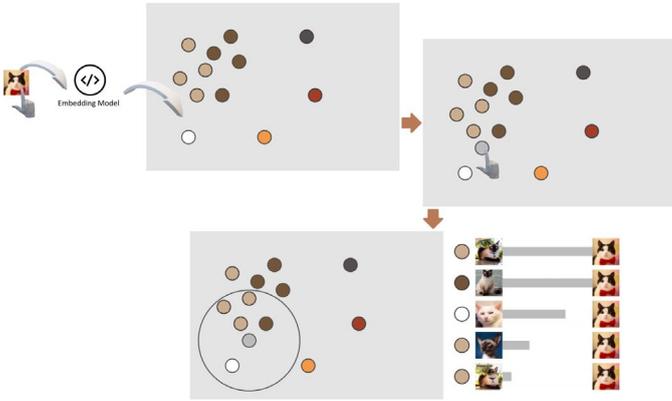

Figure 9: A new test image is fed into the embedding model, which forms a bubble in a trained embedded space to find nearby images within the new test image's radius.

#### 5.2.2. Procedure

Because collecting quantitative evaluation data is difficult due to business constraints and workplace culture, we summarize user feedback through observation and semi- and informal interviews of presenters and audiences, such as focus groups. We used scrollytelling as part of the product presentation during the demo. The proposed scrollytelling explains the backstory of this existing product, which uses SNN as the system's core.

#### 5.2.3. Feedback

The proposed scrollytelling visualization was presented to the users, and here is their feedback. For the feedback overview, the best thing about this project is that it lets people use interactive media to learn about hard ideas at their own pace. The audiences mentioned that this scrollytelling medium combines the merits of different mediums, *i.e.*, articles and videos, in which articles allow users to experience at the reader's pace. In contrast, videos break down complex ideas into more appealing visuals. Controlling the pace and narrative style of scrollytelling is critical for enabling media consumers to learn on demand. In other words, if an issue cannot be handled while watching, it may be scrolled up and down, suggesting it has control over the narrative flow.

Moreover, both audiences and the business team, as the presenters mentioned the interactions are added in each part along the way as the users scroll down, which allows users to explore the effects of these concepts in different scenarios. This is useful for reinforcing knowledge of some complex machine learning concepts. With these advantages, the scrollytelling visualization was used in business conference presentations to provide a quick summary of how the specific ML model works without wasting time explaining it to the audiences.

Also, when giving a short presentation, scrollytelling makes it easier for presenters to give their talks by letting them skim through the visualization. Moreover, during the Q&A session, the presenter can use the interactions to explain specific concepts that were asked in detail, and the audience can replay it themselves simultaneously.

#### 5.3. User Study

We conducted a statistical user study to determine how VISHIEN-MAAT's target users (i.e., business IT students) would use scrollytelling as a medium tutorial to learn about SNN in comparison to other mediums.

#### 5.3.1. Participants

This user study assesses the role of visualization in knowledge acquisition for non-technical users. Participants in the user study include 50 business IT students in the academic year 2021/2022. They enroll in the class about business intelligence and tools for data analytics, which has to learn about data science and machine learning too, at Thammasat University in Thailand. The students are expected to be able to communicate about AI/machine learning to business stakeholders and non-technical users. However, unlike engineering students, business IT students do not possess in-depth knowledge of AI/machine learning. They were divided into two groups for testing in two different mediums (scrollytelling and online articles).

#### 5.3.2. Procedure

This study will look into whether our scrollytelling visualization can help students learn the concept and application of SNN better than traditional materials. The pre-test and post-test are created to evaluate participants' understandings after learning the topics using two different mediums—scrollytelling, and online articles.

We select online articles about image similarity estimation using a Siamese Network with a contrastive loss[5] and triplet loss[6] from Keras. It is a well-known open-source software library that provides a Python interface for artificial neural networks which the topics are related to our work and are being used in class nowadays.

The online articles were chosen for comparison because they are both common choices for extra reading and have a scrolling feature like scrollytelling.

---

[5]https://keras.io/examples/vision/siamese_contrastive/
[6]https://keras.io/examples/vision/siamese_network/

With this feature, the user can control how fast the story moves by pausing or scrolling at any point while playing. This feature allows users to study the content of interest according to their needs, which is different from the other comparable genre, *e.g.*, data comics, and data-GIFs.

Seven multiple-choice questions are on the pre-test, and the same seven questions are on the post-test. The test topics include the SNN model concept, Embedding model, Loss function, Euclidean distance, Model training, and Inferencing. Each question has four multiple choices with one correct expected answer that will receive one point when answered correctly. The test questions and their correct answers as italic are listed in Appendix A, and they are given to participants in this order, with no shuffle.

The expected results of the seven questions that aim to reflect the user understanding regarding the Siamese Neural Network can be described as follows:

1. The first question can represent the conceptual architecture of the model such that Siamese words, in this case, mean Siamese Twin, which describes the model origin, which is sometimes called a twin neural network that uses the same weights while working in tandem on two different input vectors.

2. The second question needs to know that the user can understand the different loss functions used in the Siamese Neural Network depending on the architecture of the model, which is twin and triplet.

3. The third question needs the user to know the CNN, which is a core concept of each neural network in the twin and triplet architecture that suits the work for the embedded images into space.

4. The fourth question is meant to find out if the user knows the name of the input image for the triplet architecture, which can be used to point to the expected location of the embedded object in the space.

5. The fifth question relies on the previous question, that the user should observe the movement of each triplet image as it moves closer or far away.

6. The sixth question also relies on the previous question, that the user should tell us about the movement's meaning, which expects that the user knows that the similar image should be moved closer to each other than the different image.

7. The last question needs the user to tell the application of the business problem that the model-trained result can solve. For example, use the trained model to find an image similar to another by comparing all images' closeness.

The quasi-experiment was applied, and we define the process of the user study as consisting of four steps [20], as follows:

1. All participants were asked to complete the pre-test.

2. The participants were divided into two groups. They were asked to learn about the specific AI model but with different mediums. The first group uses the VISHIEN-MAAT: Siamese Neural Network Scrollytelling, while the second group studies through the two online articles from Keras, which were used as the basic course material for the students, about the SNN model with contrastive loss and triplet loss. The students need to take time in class to read the articles because they are required to read the two articles one by one with no skipping.

3. All participants are asked to complete the post-test.

4. To compare the results of two groups, perform a quantitative analysis with a t-test dependent sample by condition: $alpha = 0.05$, 2-tails.

The test instrument was used for pre-test and post-test with identical question characteristics on each test related to Siamese Neural Network. The pre-test was given before the two groups were subjected to scrollytelling for the experimental, and online articles were given for the control group. After treatment was applied to the experimental and control groups, the post-test was granted which the raw results are in Appendix B. The next step was to compare the pre-test results to the ones of the post-test for both groups.

### 5.3.3. Result

In this study, data was analyzed using Levene's test to determine variance equality. The data is equal variances assumed if Levene's test output F-test significance is more than 0.05. As an inferential statistical test, the independent sample t-Test criteria with a value of 0.05 were used. Rejection of the null hypothesis was done if the result is less than the critical value. On the contrary, failing to reject the null hypothesis was concluded if the test statistic is greater than the critical value.

The comparison of the pre-test of seven questions results between the scrollytelling and online articles mediums sorted displayed in table 2. In the comparison of the scrollytelling and online articles, the pre-test means the score is 2.92 and 3.04 respectively. While the post-test comparison between scrollytelling and online article mediums is shown in Tables 3.

Table 2: Summary table for comparing tests from different mediums

| Mediums | N | Pre test mean | Post-test mean | Std. Deviation | S.E. Mean |
|---|---|---|---|---|---|
| Scrollytelling | 26 | 2.92 | 5.85 | 1.54 | .30 |
| Online articles | 24 | 3.04 | 3.96 | 1.46 | .30 |

Table 3 displays that the F-test of post-test data with the test of equality of variances with F-test is 0.09 and sig value is 0.771 bigger than 0.05 (Sig. $> 0.05$), that is equal variances assumed between scrollytelling and online articles mediums. Then do the t-Test of post-test data at the first line of equal variances assumed with df is 48.00 in scrollytelling and online articles mediums with sig value 2-tailed is 0.000 smaller than 0.05 (Sig. $< 0.05$). In addition to this, the calculated t-values for the post-test of scrollytelling and online articles mediums achieve 4.44 more significance than the t-test value. These t-test results indicate that the null hypothesis is rejected, showing a significant difference between the post-test of experimental and control groups, which is the mean of scrollytelling (post-test mean is 5.85) is higher than the mean of the online articles (post-test mean is 3.96).

Table 3: Independent samples test of scrollytelling and online articles post-test

| | | Levene's Test for Equality of Variances | | T-Test for Equality of Means | | | | | 95% Confidence Interval of the Difference | |
|---|---|---|---|---|---|---|---|---|---|---|
| | | F | Sig. | t | df | Sig. (2-tailed) | Mean Difference | Std. Error Difference | Lower | Upper |
| Score | Equal variances assumed | .09 | .771 | 4.44 | 48.00 | .000 | 1.89 | .43 | 1.03 | 2.74 |
| | Equal variances not assumed | | | 4.45 | 47.97 | .000 | 1.89 | .42 | 1.03 | 2.74 |

## 6. Discussion

Based on the visual similarity matching problem context in real business, we designed the scrollytelling visualization of the SNN model for business presentations. The above goal can be expanded to include a learning test with business students to compare to another medium.

In the business presentation scenario, the audience and presenters used the scrolling features to control the story's pace and the interactive parts to freely explore the steps of the model during the sales meeting. The user study results show that the scrollytelling visualization design is a better way to support the achievement of learning SNN model concepts than online articles. The students may understand the ML concept, especially SNN, more efficiently with the scrollytelling visualization design because they have complete control of the pace. They can go over complex or new ideas again if needed by scrolling up or down to pause, play, rewind, or even fast-forward the material on demand. Moreover, users can also participate in the learning via interactive components, where they can explore what will happen in different steps of the SNN.

We discuss how scrollytelling can help close gaps in online articles, video tutorials, data comics, and data-GIFs. The online articles are unappealing, causing the user to lose focus on the content. Our scrollytelling makes use of interaction and animation to keep users' attention. The disadvantage of video tutorials is that users cannot control the narrative's pace. However, with scrollytelling, users can scroll up and down to meet their needs. Because no interactions are available, data comics have the same limitations as online articles. The main advantage of data comics over online articles is that they appeal to the reader; scrollytelling, on the other hand, uses interaction and animation to help catch the attention of users. In some ways, the fact that data-GIFs can make an animation between the content and themselves makes them easier to understand. However, there are time constraints on when data-GIFs can be used. Scrollytelling, on the other hand, is open to interpretation as long as the creators of the content determine what is appropriate.

The complexity of the deep learning concept that needs to be communicated is the limitation of the scrollytelling design. When the stories are complicated, the scrollytelling will be longer. Rich interactive experiences can be overwhelming, so the length of the stories should be limited to keep the effectiveness of the visual storytelling.

## 7. Conclusion and Future Work

This study presents a novel visualization design to help comprehend the Siamese Neural Network (SNN) model, which is easier to realize than traditional mediums for non-technical users. The interactive scrollytelling visualization demonstrates that the proposed approach effectively explains the similarity matching model used in the industry and communicates the model's mechanism for business and education purposes. With interactive scrollytelling, users can look into certain points at their own pace, and the interactive parts break down difficult information into pieces that are easy to understand. Both quantitative and qualitative evaluations suggest that our method helps non-technical end-users, *i.e.*, the business marketing team and the business IT students, quickly understand the deep learning model. The design for explaining the SNN model allows for advanced techniques such as 3D interaction components, which is a good opportunity for future work in which we can apply this scrollytelling design and add the step of rich details and rich interactions in each story slice.

**Acknowledgement**

This paper is partially supported by the National Natural Science Foundation of China (No. 62132017). The first author wishes to thank Mr. Wissarut Pimanmassuriya, Ms. Panita Rerkpitivit, Ms. Chalida Liamwiset and Mr. Kittikun Kamrai for their valuable technical, data support on this project.

## Appendix A. Pre-post Tests' Questions and Answers

The seven pre-post tests' questions and answers for online articles and purposed scrollytelling.

1. What is the origin of the word Siamese in Siamese Neural Network? The answer is 'Siamese Twin'.

2. Which loss function is associated with the term Siamese? The answer is 'Contrastive Loss'.

3. What kind of network is the most popular at the heart of the image processing Embedding Model? The answer is 'CNN or ConvNet'.

4. A batch of data is organized as follows: 1) The original image 2) The image is nearly identical to the original. 3) The image does not resemble the original. What is the name of this type of batch? The answer is 'Triplets'.

5. What should you expect to observe throughout training? The answer is 'Loss value gradually decreases'.

6. What does it mean if similar objects are close together and dissimilar objects are separated? The answer is 'Loss value gradually decreases'.

7. How will the final trained model be able to solve the business problem? The answer is the combination of things the SNN-trained model can do, in which it can compare a new object to a previously trained surrounding object, then find the new object, and the trained model can be used as a pre-trained model.

## Appendix B. Raw Results

The raw results of pre-post tests are sorted by participant ID of online articles and purposed scrollytelling.

Table B.4: The raw results of the online article and scrollytelling pre-post tests

| Online article | | | | Scrollytelling | | | |
|---|---|---|---|---|---|---|---|
| PID | Pre | Post | Post-Pre Diff. | PID | Pre | Post | Post-Pre Diff. |
| 1 | 2 | 3 | 1 | 1 | 2 | 7 | 5 |
| 2 | 4 | 3 | -1 | 2 | 1 | 5 | 4 |
| 3 | 4 | 5 | 1 | 3 | 3 | 7 | 4 |
| 4 | 4 | 4 | 0 | 4 | 3 | 7 | 4 |
| 5 | 4 | 3 | -1 | 5 | 3 | 5 | 2 |
| 6 | 2 | 3 | 1 | 6 | 4 | 3 | -1 |
| 7 | 3 | 4 | 1 | 7 | 3 | 6 | 3 |
| 8 | 3 | 4 | 1 | 8 | 2 | 7 | 5 |
| 9 | 4 | 3 | -1 | 9 | 4 | 6 | 2 |
| 10 | 3 | 3 | 0 | 10 | 3 | 2 | -1 |
| 11 | 3 | 3 | 0 | 11 | 4 | 7 | 3 |
| 12 | 3 | 3 | 0 | 12 | 1 | 7 | 6 |
| 13 | 5 | 3 | -2 | 13 | 1 | 7 | 6 |
| 14 | 3 | 2 | -1 | 14 | 5 | 7 | 2 |
| 15 | 4 | 4 | 0 | 15 | 3 | 7 | 4 |
| 16 | 2 | 4 | 2 | 16 | 3 | 5 | 2 |
| 17 | 4 | 3 | -1 | 17 | 4 | 5 | 1 |
| 18 | 4 | 2 | -2 | 18 | 4 | 5 | 1 |
| 19 | 4 | 6 | 2 | 19 | 3 | 7 | 4 |
| 20 | 1 | 7 | 6 | 20 | 4 | 5 | 1 |
| 21 | 0 | 6 | 6 | 21 | 3 | 7 | 4 |
| 22 | 3 | 7 | 4 | 22 | 5 | 7 | 2 |
| 23 | 1 | 6 | 5 | 23 | 2 | 6 | 4 |
| 24 | 3 | 4 | 1 | 24 | 2 | 7 | 5 |
| | | | | 25 | 2 | 2 | 0 |
| | | | | 26 | 2 | 6 | 4 |